\documentclass[aps,prl,twocolumn,reprint]{revtex4-2}
\usepackage{graphicx}
\usepackage{amsmath, amssymb, amsfonts}
\usepackage{xcolor}
\usepackage[colorlinks=true,linkcolor=blue]{hyperref}
\usepackage[normalem]{ulem}
\usepackage{tikz}

\usepackage{pdfpages}
\usepackage{pgffor}

\makeatletter
\AtBeginDocument{\let\LS@rot\@undefined}
\makeatother

\def\supplementfilename{Shielding-mw-onsite-letter-supp-resub.pdf}

\pdfximage{\supplementfilename}
\def\numbersupplementpages{\the\pdflastximagepages}

\newif\ifarXiv
\arXivtrue

\begin{document}
\title{On-site interaction energies for Hubbard models with shielded polar molecules}

\author{Carlin Stewart-Wiese}
\affiliation{Joint Quantum Centre (JQC)
Durham-Newcastle, Department of Chemistry, Durham University, South Road,
Durham, DH1 3LE, United Kingdom.}

\author{Joy Dutta}
\affiliation{Joint Quantum Centre (JQC)
Durham-Newcastle, Department of Chemistry, Durham University, South Road,
Durham, DH1 3LE, United Kingdom.}

\author{Jeremy M. Hutson}
{\email{j.m.hutson@durham.ac.uk} \affiliation{Joint Quantum Centre (JQC)
Durham-Newcastle, Department of Chemistry, Durham University, South Road,
Durham, DH1 3LE, United Kingdom.}

\date{\today}

\begin{abstract}
We show how to control on-site interaction energies $U$ for pairs of shielded ultracold molecules on the same lattice site. We use 2-dimensional effective potentials appropriate for microwave shielding, which have dipole-dipole character at long range but feature a very large repulsive core when the two molecules come close together. This causes very strong correlation between the motions of two molecules on the same site and produces field-linked states. We identify field-linked states for which $U$ can be tuned over a wide range, including across zero, as a function of lattice strength or shielding conditions. This will give access to previously unexplored regimes of strongly dipolar Hubbard physics with multiple site occupancy.
\end{abstract}

\maketitle

\section{Introduction}
Ultracold polar molecules provide an important platform for quantum science. They have properties distinct from ultracold atoms, because they have long-range dipolar interactions that can be controlled with applied electric and microwave fields. They have emerging applications in quantum simulation and quantum computation \cite{Cornish:2024}, many-body physics \cite{Baranov:2012, Bohn:q-eng:2017, Schindewolf:few-many:2025}, quantum magnetism \cite{Micheli:2006, Wall:QMUM:2015}, quantum metrology \cite{DeMille:2024}, and controlled chemical reactions \cite{Karman:2024}. In the last few years, both Fermi degeneracy \cite{Schindewolf:NaK-degen:2022} and Bose-Einstein condensation (BEC) \cite{Bigagli:BEC:NaCs:2024, Shi:BEC:NaRb:2026} have been achieved for gases of ultracold molecules.

Many interesting properties arise for ultracold molecules confined in optical lattices \cite{Cornish:2024}. These properties can loosely be divided into those for molecules that are pinned on lattice sites \cite{Micheli:2006, Yan:2013, Wall:magnets:2013, Moses:2015, Wall:2015, Wall:QMUM:2015, Christakis:2023, Mortlock:2026} and those where the molecules can tunnel between sites \cite{Barnett:2006, Wall:2010, Gorshkov:PRA:2011, Kestner:2011, Wall:2013, Docaj:2016, Wall:2017}. The present paper focusses on the latter case, which we refer to generically as Hubbard physics. Recent experimental work with fermionic KRb molecules has explored the coherent spin dynamics of molecules tunneling in 1 and 2 dimensions \cite{Carroll:t-J:2025}. Analogous experiments with bosonic molecules are likely to be possible soon.

Most experimental and theoretical work so far has excluded the possibility of two molecules occupying the same site \cite{Wall:2010, Gorshkov:PRA:2011, Kestner:2011, Wall:2013, Carroll:t-J:2025}. However, many-body systems show rich physics when multiple occupancy is allowed, particularly when there are strong interactions between particles on different sites \cite{Dutta:2015}. Effects such as pair superfluidity emerge \cite{Daley:2009}, together with exotic solid and supersolid phases \cite{vanOtterlo:1995} and a wide variety of quantum phase transitions. The general form of the Hamiltonian for Hubbard models is described below. The phases and phase transitions are principally governed by the competition between the tunneling energy $t$ and on-site interaction energy $U$, together with additional parameters that characterize longer-range interactions and occupation-dependent effects \cite{Dutta:occ:2011}. Molecules provide an outstanding platform for exploring these phases and effects, because their large dipole moments provide strong interactions between molecules on nearby sites.

Ultracold molecules behave differently from both magnetic and non-magnetic atoms. In particular, pairs of molecules that occupy the same lattice site (doublons) will usually undergo fast collisional loss. To avoid this, it will usually be necessary to use shielding techniques based on either static electric \cite{Avdeenkov:2006, Wang:dipolar:2015, Gonzalez-Martinez:adim:2017, Mukherjee:CaF:2023, Mukherjee:alkali:2024} or microwave \cite{Karman:shielding:2018, Lassabliere:2018, Karman:shielding-imp:2019, Karman:ellip:2020, Karman:res:2022, Deng:microwave:2023, Dutta:universal:2025, Karman:double:2025} fields. These techniques suppress the loss by creating repulsive barriers at separations $R$ that are typically 1000 bohr. The repulsive barriers prevent access to the dense manifold of short-range two-molecule states \cite{Mayle:2013} that produce multichannel interactions for unshielded molecules in lattices \cite{Docaj:2016, Wall:2017}. The corresponding effective potentials \cite{Deng:microwave:2023, Deng:double-microwave:2025, Karman:double:2025, Mukherjee:eff-pot:2025} have large excluded volumes at small $R$ and strong dipole-dipole interactions at large $R$. Both of these strongly modify the simple dependence of $U$ on scattering length and harmonic frequency that is found for atoms \cite{Jaksch:1998}. The purpose of the present paper is to explore the behavior of $U$ for shielded molecules, using realistic interaction potentials. Understanding the tunability of $U$ is crucial to realizing the full potential of ultracold molecules in many-body physics.

\section{Methods}

The simplest Hubbard model is the single-band Bose-Hubbard model, with Hamiltonian
\begin{equation}
\hat{H}_\textrm{BH} = -t \sum_{ij'} a^\dagger_i a_j + \frac{1}{2} U \sum_i n_i(n_i-1),
\end{equation}
where $a^\dagger_i$ and $a_j$ are bosonic creation and annihilation operators for particles on sites $i$ and $j$, respectively, and $n_i$ is the number of particles on site $i$. The sum over $i$ and $j$ is over nearest-neighbor pairs.

Many other Hubbard models have been studied for both bosons and fermions \cite{Dutta:2015}, including extensions to account for interactions between particles on different sites, occupation-dependent parameters, superexchange, nonadditive effects, etc. Models that include long-range dipolar interactions are particularly important \cite{Baranov:2012}. The calculations of onsite interaction energies $U$ in the present paper may be used in all these models.

The usual way of calculating $U$ is based on Wannier functions, which are real-space functions for independent particles, expressed with respect to the center of each lattice site \cite{Ashcroft:solid:1976}. For two particles whose motions are not strongly correlated, $U$ may be approximated as an integral of the interaction potential $V(\boldsymbol{r}_1,\boldsymbol{r}_2)$ over Wannier functions for the lowest band \cite{Jaksch:1998}.
For ultracold atoms, $V(\boldsymbol{r}_1,\boldsymbol{r}_2)$ is often approximated as a regularized contact potential \cite{Huang:1957} characterized by the scattering length $a$. Jaksch \emph{et al.}\ \cite{Jaksch:1998} showed that, for a spherical lattice cell, this yields for $a \ll \beta$
\begin{equation}
U \approx \sqrt{\frac{2}{\pi}} \hbar\omega (a/\beta),
\label{eq:Jaksch}
\end{equation}
where $\omega$ is the harmonic frequency and $\beta$ is the single-particle harmonic length.

For dipolar particles, the contact potential is often supplemented by a long-range dipole-dipole interaction \cite{Yi:2000, Yi:2001}. When $a$ is relatively small and the dipole-dipole interaction is weak, correlations between the motions of two particles on the same site may be neglected. In this approximation, the dipolar and contact contributions to $U$ are additive \cite{Goral:latt-dip:2002, Sowinski:2012} and may be adjusted independently \cite{Menotti:2008}. The dipolar interaction averages to zero for two particles on a spherical site.

The approach above has been successful for high-spin atoms \cite{Chomaz:2023}, where the dipolar interactions are relatively weak. However, neglecting on-site correlations is a poor approximation for shielded molecules. The repulsive core of the shielding potential produces a large excluded volume around $\boldsymbol{R}=\boldsymbol{r}_1-\boldsymbol{r}_2=0$. At the same time, the dipolar interactions may be strong enough to cause $R$-dependent angular localization, so that the dipolar interaction does not average to zero even on a spherical site. The motions of two molecules on the same site are thus strongly correlated, and are not well represented by any simple product of functions of $\boldsymbol{r}_1$ and $\boldsymbol{r}_2$.} Under these circumstances, an approach based on Wannier functions for a single band is inappropriate.

In this paper we use a different approach that takes full account of on-site correlation. For two particles on the same site, we separate the relative and center-of mass motions and approximate the confining potential as a single harmonic well. This approach allows exact numerical solution of the correlated 2-particle problem, because the relative and center-of-mass motions are exactly separable for two equally trapped particles in a harmonic well \cite{Deuretzbacher:2008}.

The Schr\"odinger equation for relative motion is
\begin{align}
\bigg[\frac{-\hbar^2}{2\mu_\textrm{red}} &\left(\frac{1}{R} \frac{d^2}{dR^2} R \right) +
\frac{\hbar^2}{2\mu_\textrm{red}} \frac{\hat{\boldsymbol{L}}^2}{R^2} \nonumber\\
&+ V(\boldsymbol{R})+ V_\textrm{confine}(R) - E_\textrm{rel}\bigg]\Psi(\boldsymbol{R})=0.
\label{eq:schrod_rel}
\end{align}
Here the first and second terms represent the kinetic energy for relative radial and rotational motion; $\mu_\textrm{red}$ is the reduced mass, and $\hat{\boldsymbol{L}}$ is the operator for relative angular momentum. The effective potential $V(\boldsymbol{R})$ is described below, and the confining potential is $V_\textrm{confine}(R)=\frac{1}{2}\mu_\textrm{red}\omega^2 R^2$.

The total energy of the pair is $E_\textrm{tot} = E_\textrm{rel} + E_\textrm{com}$, where $E_\textrm{rel}$ is the energy of relative motion and the center-of-mass energy $E_\textrm{com}$ is the same for 2 particles on the same or different sites. The relative energy of 2 particles on separate sites (in 3 dimensions) is just $\frac{3}{2}\hbar\omega$, so
\begin{equation}
U = E_\textrm{rel} - \textstyle{\frac{3}{2}} \hbar\omega.
\end{equation}
In the present work, we obtain $E_\textrm{rel}$ from numerically exact coupled-channel calculations.
We expand the wavefunction $\Psi(\boldsymbol{R})$ in a coupled-channel representation as described in Supplemental Material \cite{Stewart-Wiese:2026:SM}.~\nocite{Hutson:CPC:1994,Manolopoulos:1986,Alexander:1987}
The resulting coupled equations have the same structure as those for pure dipolar scattering \cite{Bohn:BCT:2009} or bound states \cite{Karman:dipole:2018}, but with a different interaction potential. We solve them using the BOUND package \cite{bound+field:2019, mbf-github:2025}, which applies bound-state boundary conditions at long and short range and propagates incoming and outgoing coupled-channel solutions to a matching point in the classically allowed region. Coupled-channel scattering calculations without the confining potential $V_\textrm{confine}(R)$ allow us to calculate the corresponding scattering length $a$ as described in ref.\ \cite{Hutson:res:2007} and are carried out with the MOLSCAT package \cite{molscat:2019, mbf-github:2025}. Numerical details for both bound and scattering calculations are included in Supplemental Material \cite{Stewart-Wiese:2026:SM}.

Our approach neglects the anharmonicity of the confining potential, including any extension of the Wannier functions into neighboring wells. Nevertheless, it is expected to be reasonably accurate for lattice strengths $V_0 \gtrsim 5E_\textrm{r}$, where $E_\textrm{r}$ is the recoil energy. Lattices of at least this strength are in any case necessary to implement single-band Hubbard models, which require a significant band gap between the lowest band and higher ones.

For studies of Hubbard models, it is natural to compare $U$ with the tunneling energy $t$ between adjacent lattice sites. In the simplest approach, $t$ is a one-particle property that can be calculated using standard methods \cite{Blakie:2004}. It decreases very fast as the lattice strength $V_0$ increases. The Mott-insulator transition for non-polar bosons typically occurs around $U/t \sim 5.8z$ \cite{Sheshadri:1993, Greiner:2002}, where $z$ is the number of nearest neighbors. Bose-Hubbard models with and without nearest-neighbor interactions show rich phase behavior at smaller values of $U/t$ \cite{Kuhner:1998}. Attractive Fermi-Hubbard and Bose-Hubbard models, with negative $U$, are also of interest \cite{Dutta:2015, Jack:2005}. In general terms, interesting properties involving multiple site occupancy occur for values of $U/t$ between about $-10$ and 50.

In this paper we focus on pairs of molecules shielded with circularly polarized microwaves \cite{Karman:shielding:2018, Lassabliere:2018, Deng:microwave:2023, Dutta:universal:2025}. We use the 2-dimensional (2d) effective potentials $V(\boldsymbol{R})=V(R,\theta)$ of ref.\ \cite{Deng:microwave:2023}, which depend on the intermolecular distance $R$ and the angle $\theta$ between the intermolecular vector and the axis of polarization. Analogous effective potentials have been obtained for static-field shielding \cite{Mukherjee:eff-pot:2025}, double-field microwave shielding \cite{Karman:double:2025, Deng:double-microwave:2025}, and single-field shielding with elliptical polarization \cite{Deng:microwave:2023}. Extension to these other cases will be possible in future.

\section{Results and Discussion}

To provide a definite example, we carry out calculations on NaCs molecules shielded by circularly polarized microwaves with Rabi frequency $\Omega$, blue-detuned by $\Delta$ from the rotational transition $n=0 \rightarrow 1$. However, our results for bound-state and on-site interaction energies are universal in the sense described in ref.\ \cite{Dutta:universal:2025}: they would be the same for any polar molecule if all energies ($E_\textrm{rel}$ and $U$) and frequencies ($\Omega$, $\Delta$ and $\omega$) are expressed in terms of the molecule-dependent quantity $E_3=\hbar^2/(2\mu_\textrm{red}R_3^2)$, where $R_3=(2\mu_\textrm{red}/\hbar^2)(\mu^2/4\pi\epsilon_0)$, $\mu$ is the molecular dipole moment, and $\mu_\textrm{red}$ is the 2-molecule reduced mass. However, the recoil energy $E_\textrm{r}$ is independent of $\mu$, so that $U/t$ is non-universal.

For ultracold atoms, $U$ typically depends on the scattering length $a$ and the harmonic frequency $\omega$ according to Eq.\ \ref{eq:Jaksch}. It is a monotonic function of $\omega$ and its sign depends on the sign of $a$. For shielded molecules, $a$ can be adjusted by varying the shielding fields. For some shielding conditions, the effective potential supports weakly bound states even in the absence of confinement, and $a$ passes through poles where they appear. For NaCs with $\Delta/\Omega=1$, poles appear at $\Omega=0.77$ and 13.2 MHz. Everywhere between these two poles, the effective potential is deep enough to support one bound state.

\begin{figure}[tb]
\centering
\includegraphics[width=1\columnwidth]{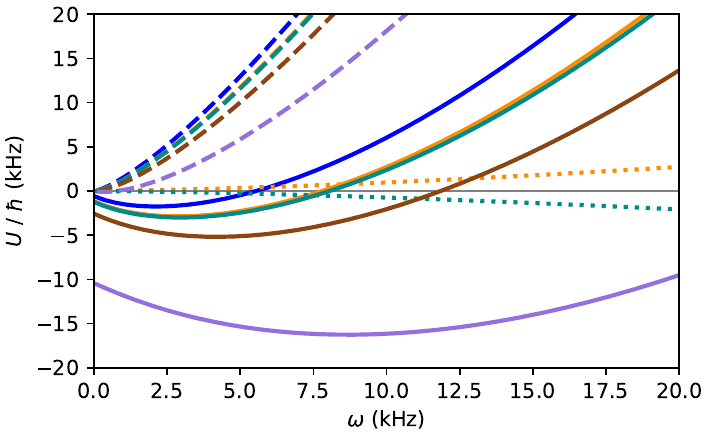}
\caption{On-site interaction energy $U$ as a function of harmonic frequency $\omega$, for NaCs molecules shielded with circularly polarized microwaves with $\Delta/\Omega=1$. States that are bound (unbound) at $\omega=0$ are shown as solid (dashed) lines. Rabi frequencies $\Omega=2.5$ (blue), 3.4 (orange), 3.5 (green), 5 (brown) and 11 MHz (purple) give scattering lengths $a=5200$, 260, $-200$, $-6400$ and $-100,000$ bohr, respectively. Dotted lines show the results of Eq.\ \ref{eq:Jaksch} for the lowest two values of $|a|$. The value of $U$ for the bound states (field-linked states) can be tuned across a wide range by varying either harmonic frequency or microwave parameters. The results are universal; values for other polar molecules may be obtained by scaling all frequencies by $E_3$ as described in the text.}
\label{fig:U-single}
\end{figure}

Figure \ref{fig:U-single} shows the results of coupled-channel calculations of $U(\omega)$ at several values of $\Omega$, chosen to span the range between moderate positive and large negative scattering lengths. For each value of $\Omega$, there are two curves: the lower curve is for a two-molecule state that is bound even at $\omega=0$, while the upper curve is for a state that is unbound at $\omega=0$; we refer to these below as the bound and unbound states. The bound states are sometimes termed field-linked states \cite{Avdeenkov:2003, Avdeenkov:2004}.

We first consider the unbound state, which is the closest analog of the state usually used for Hubbard physics with ultracold atoms. For shielded molecules with positive $a$, $U(\omega)$ for this state increases monotonically with $\omega$. For negative $a$, it dips below zero at small $\omega$, but crosses zero to become positive well below $\omega=2$ kHz. Even for small scattering lengths, $U(\omega)$ differs greatly from the results of Eq.\ \ref{eq:Jaksch}, which are shown as dashed lines in Fig.\ \ref{fig:U-single} for the smallest positive and negative values of $a$. The differences arises mostly from the large repulsive core of the shielding potential: the wavefunction is effectively squeezed between the repulsive core and the confining potential, raising its energy above that expected for a contact potential with the same scattering length.

For NaCs molecules in a lattice with $\lambda = 1064$ nm, the recoil energy is $E_\textrm{r}/\hbar = 1.13$ kHz. The minimum lattice strength $V_0=5E_\textrm{r}$ that meets the requirement of single-band Hubbard models thus corresponds to $\omega\approx 5$ kHz. For all the cases in Fig.\ \ref{fig:U-single}, $U(\omega)$ for the unbound state is at least 5 kHz at $\omega=5$ kHz. At $V_0=5E_\textrm{r}$, $t=0.066E_\textrm{r}=0.075$ kHz, so $U > 5$ kHz corresponds to $U/t > 60$. Such large values of $U/t$ are likely to prevent multiple occupation of lattice sites, effectively applying a hard-core constraint to the resulting Hubbard physics.

The bound states in Fig.\ \ref{fig:U-single} show quite different behavior. Their binding energies in the absence of confinement are much smaller than for the corresponding states for atoms, which seldom have binding energies less than 50 kHz \cite{Berninger:Cs2:2013}. As a confining potential is applied, $E_\textrm{rel}$ initially increases quadratically with $\omega$. However, the energy of two molecules on separate sites is $\frac{3}{2}\hbar\omega$, so increases faster. As a result, $U$ initially decreases (becomes more negative) at low $\omega$, with the same initial gradient $-\frac{3}{2}\hbar$ for all $\Omega$. At higher $\omega$, however, squeezing of the wavefunction between the core of the shielding potential and the confining potential again dominates; as a result, $U$ increases and crosses zero at finite $\omega$. The value of $\omega$ at the crossing point increases with the initial binding energy, but in many cases occurs in an experimentally accessible range of lattice strength. Contour plots of wavefunction densities for representative confined states are shown in Supplemental Material \cite{Stewart-Wiese:2026:SM}.

\begin{figure}[tb]
\centering
\includegraphics[width=1\columnwidth]{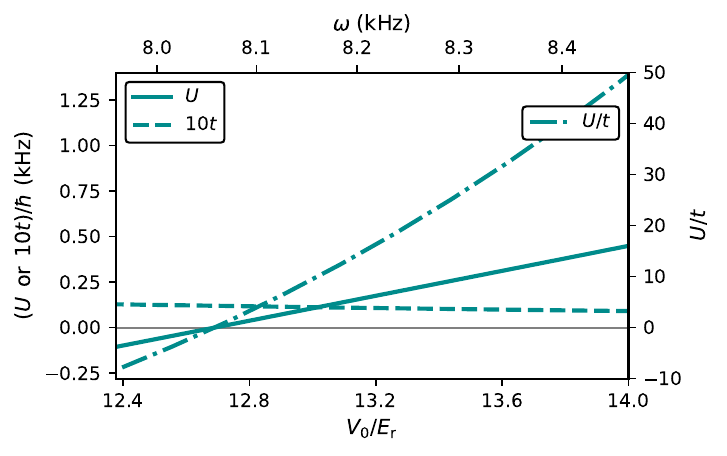}
\caption{On-site interaction energy $U$ (solid, left axis), tunneling energy $t$ (dashed, left axis) and their ratio $U/t$ (dot-dashed, right axis), as a function of lattice strength $V_0$ and harmonic frequency $\omega$, near the zero-crossing of $U$. The calculations are for NaCs with $\Omega=3.5$ MHz, $\Delta/\Omega=1$ and $\lambda=1064$ nm. The value of $U/t$ can be tuned across the full range of interest for attractive and repulsive Hubbard models.}
\label{fig:U-t}
\end{figure}

Figure \ref{fig:U-t} shows $U$, $t$ and $U/t$ as a function of lattice strength for one of the cases in Fig.\ \ref{fig:U-single}, with $\Omega=3.5$~MHz and $\Delta/\Omega=1$. The corresponding harmonic frequency is shown on the upper axis. It may be seen that $U/t$ is easily tuned from $-10$ to $+50$ simply by varying the lattice strength. Alternatively, it may be tuned by varying the Rabi frequency $\Omega$, which shifts the zero-crossing to a different value of $\omega$ as shown in Fig.\ \ref{fig:U-single}. The two-molecule bound states thus offer ready tuning of $U$ across the entire range of $U/t$ that is needed for studies of Hubbard models \emph{without} a hard-core constraint.

In the gas phase, three-body recombination may cause loss whenever a two-body bound state exists \cite{Bigagli:BEC:NaCs:2024}. It involves formation of a dimer and a monomer from three monomers, with release of kinetic energy. In a lattice, the corresponding process is formation of a doublon (of relative energy $U$), with any excess energy causing excitation into a higher band (typical energy $\hbar\omega$). It will thus be unimportant in the single-band regime, when $U + \hbar\omega \gg t$.

As mentioned above, one particular reason for interest in ultracold molecules in optical lattices is that they are dipolar, with anisotropic long-range forces between molecules. Because of this, they offer access to extended Hubbard models that include interactions between molecules on different sites \cite{Dutta:2015}. For microwave-shielded NaCs with $\Delta/\Omega=1$, the effective dipole moment $d_{\rm eff}$ is $\mu/\sqrt{24} = 0.97$~D, and the dipole-dipole interaction between molecules on adjacent sites in the $xy$ plane is 0.94~kHz for $\lambda=1064$~nm, so is comparable to $U$ on the scale of Fig.\ \ref{fig:U-t}. Shielded molecules may also be used to implement many-body quantum systems with SU($N$) symmetry for both bosons and fermions and for both attractive and repulsive interactions \cite{Mukherjee:SU(N):2025, Mukherjee:SU(N)-alkali:2025}.

\section{Conclusions}

Shielding techniques allow two ultracold polar molecules to collide without destructive collisions. Pairs of shielded molecules will be able to occupy the same lattice site without fast loss. This capability will allow experimental study of a wide range of models in many-body lattice physics, without a hard-core constraint that prevents double occupancy. The large dipole moments available for polar molecules can produce much stronger dipole-dipole forces than are accessible with high-spin atoms, and the dipole moments can be tuned across a wide range by adjusting the shielding fields.

The critical extra quantity that governs double occupancy is the on-site interaction energy $U$. We have explored the behavior of $U$ for pairs of shielded molecules, as a function of both harmonic frequency $\omega$ and shielding parameters. Our calculations are for molecules shielded with microwave fields, but our general conclusions also apply to other types of shielding. For shielded molecules, both motional and angular on-site correlations are critical in determining $U$. The resulting behavior is very different from that for pairs of atoms, including high-spin atoms with magnetic dipoles. In particular, $U$ is often a non-monotonic function of $\omega$: even when it is negative for small $\omega$, it crosses zero to positive values at higher values.

For molecular pairs that are unbound at $\omega=0$, $U$ may be negative at small $\omega$ but quickly crosses zero to positive values, even when the scattering length is negative. This occurs because shielded molecules have interaction potentials with large repulsive cores, which do not occur for atoms. By the time the lattice is strong enough for studies of single-band Hubbard models, $U$ is usually large enough to prevent multiple occupancy of sites.

The interactions between shielded molecules can also support bound states (field-linked states), whose binding energies at $\omega=0$ are typically a few kHz. For these states, $U(\omega)$ initially decreases (becomes more negative), but then increases again in the same way as for unbound states. The zero-crossing often occurs at experimentally accessible harmonic frequencies, and varies with shielding parameters. The field-linked states will allow precise control of $U$ over a wide range of values, including across zero.

Models of many-body physics that allow multiple occupancy offer access to a much larger Hilbert space than those that do not. Many exotic phases and effects emerge, including pair superfluidity and a wide variety of solid and supersolid phases. Understanding and controlling the on-site interactions for shielded molecules opens the way for experimental study of all these phases, and transitions between them, in the presence of the strong dipolar forces characteristic of molecules.

\section*{Rights retention statement}

For the purpose of open access, the authors have applied a Creative Commons Attribution (CC BY) licence to any Author Accepted Manuscript version arising from this submission.

\section*{Data availability statement}

The data that support the findings of this article are openly available from Durham University at http://doi.org/10.15128/r1f1881m027.

\section*{Acknowledgement}
We are grateful to Bijit Mukherjee, Jonathan Mortlock, Aileen Durst, Sohail Dasgupta, Holly Middleton-Spencer, Hannah Price, Joe Bhaseen and Ruth Le Sueur for valuable discussions. This work was supported by the U.K. Engineering and Physical Sciences Research Council (EPSRC)
Grant Nos.\
EP/P01058X/1, 
EP/W00299X/1, 
EP/Z535898/1, 
and UKRI2226. 


\bibliographystyle{long_bib}
\bibliography{all}


\ifarXiv

    \foreach \x in {1,...,\numbersupplementpages}
    {
        \clearpage
        \includepdf[
            pages={\x},
            pagecommand={}
        ]{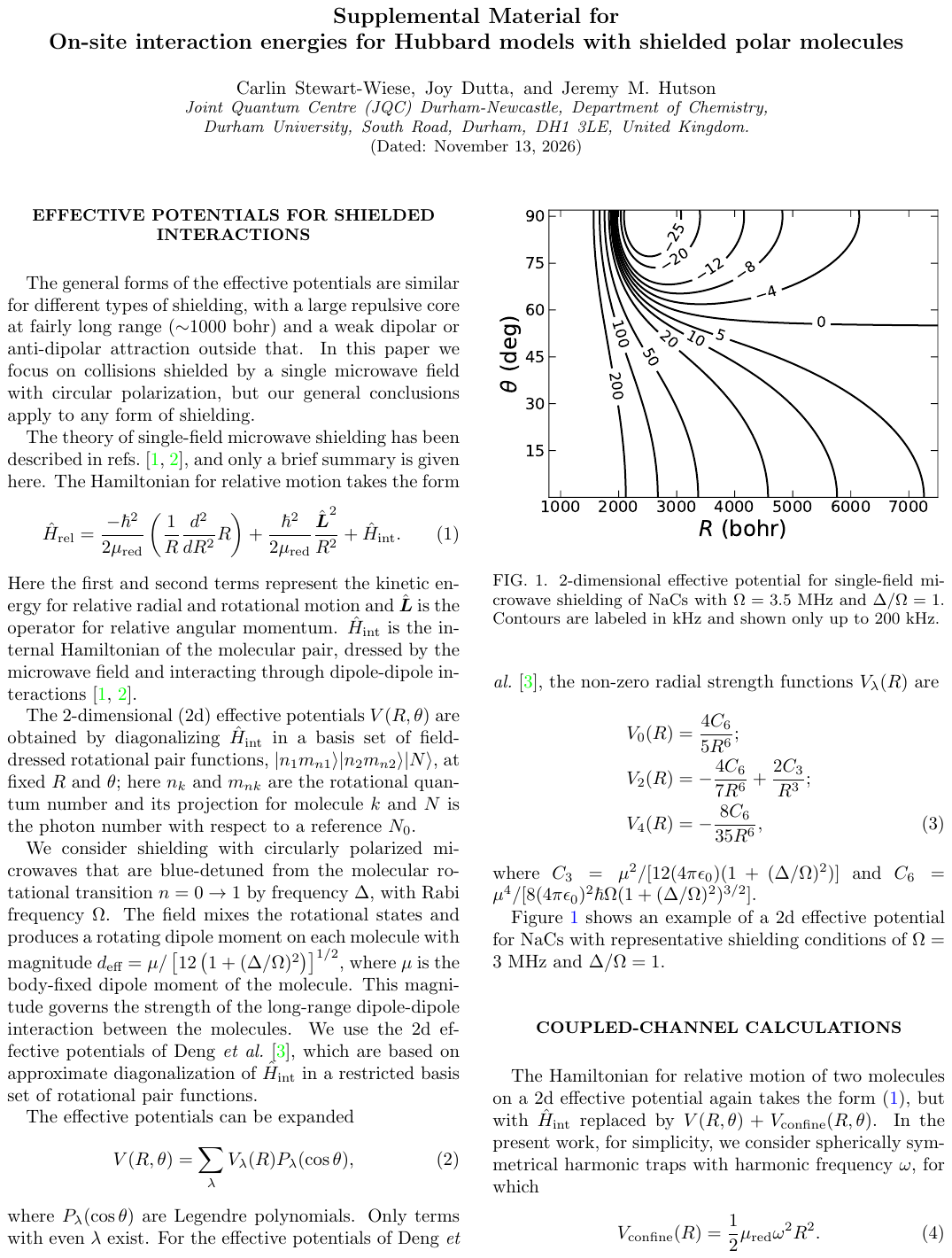}
    }

\fi

\end{document}